# New fractional classifications of papers based on two generations of references and on the ASJC Scopus scheme

**Jesús M. Álvarez Llorente[1], Vicente P. Guerrero-Bote[2], Félix de Moya-Anegón[3]**

[1] Departamento de Ingeniería de Sistemas Informáticos y Telemáticos, Universidad de Extremadura, Badajoz, Spain, llorente@unex.es, ORCID 0000-0002-4901-3457

[2] Departamento de Información and Comunicación, Universidad de Extremadura, Badajoz, Spain, guerrero@unex.es, ORCID 0000-0003-4821-9768

[3] SCImago Research Group, Granada, Spain, felix.moya@scimago.es, ORCID 0000-0002-0255-8628

## Abstract
This paper presents and evaluates a set of methods to classify individual Scopus publications using their references back to the second generation, where each publication can be assigned fractionally into up to five ASJC (All Science Journal Classifications) categories, excluding the Multidisciplinary area and the miscellaneous categories. Based on proposals by Glänzel et al. (1999a, 1999b, 2021), some additional parameters are established that allow different results to be obtained depending on how category membership is weighted or how the acceptance thresholds for multiple assignments are established. Various classifications are obtained, and then compared with each other, with the original ASJC Scopus journal classification, and with the AAC (Author's Assignation Collection) classification of a previous study (Álvarez-Llorente et al., 2023) in which the papers' corresponding authors assign them the most appropriate categories. Classifications in which a high threshold is set for allowing assignments to multiple categories, combined with the use of first- and second-generation references and averaging over the number of references, provide the most promising results, improving over other reference-based reclassification proposals in terms of granularity, and over the Scopus classification itself in such aspects as the homogeneity of the publications assigned to a category. They also show greater coincidence with the AAC classification.

## Funding
Grant project PID2020-115798RB-I00 funded by Ministerio de Ciencia e Innovación of Spain (Micin), Agencia Estatal de Investigación (AEI) / 10.13039/501100011033.

## Statements and Declarations
**Competing Interests:** The authors have no competing interests to declare that are relevant to the content of this article.

## Introduction

The assignment of papers to scientific disciplines is very important in scientometrics, not only for the disaggregation of scientific output by categories but also for the normalization of citation indicators (Althouse et al., 2009; Lancho-Barrantes et al., 2010), and in interdisciplinary studies, inter alia.

Some time ago now, Glänzel et al. (1999a, 1999b) proposed the use of papers' references in assigning them to categories, the proposal being for papers published in journals classified within the WoS's Multidisciplinary category. More recently, Glänzel et al. (2021) set out an improvement on that method with the use of a parametric model taking multiple generations of references into account.

Janssens et al. (2008, 2009) and Boyack et al. (2013) propose hybrid systems supported by text mining in addition to citation links.  One can also find studies such as that of Glenisson et al. (2005) who combine citations and text mining, although in this case applying the latter to the titles of the references instead of exploring the network they map out.  A contrary perspective had been taken in Janssens et al. (2006) who proposed classifications based on full text mining, but concluded that a combination with traditional citation-based bibliometric methods may well be appropriate.

Neural network algorithms, which since their inception have shown a great capacity to detect the key characteristics in document organization systems (Guerrero-Bote et al., 2002), have also been used as support for classification systems based on texts of titles and/or abstracts, whether the learning is supervised (Eykens et al., 2019) or unsupervised (Kandimalla et al., 2021).  As affirmed in Kandimalla et al. (2021), most unsupervised systems base their self-learning on citation networks.

Milojević (2020) uses the sum of the categories of the cited journals to assign a single category to all WoS items (those that have some classifiable reference), and add an iterative pass to increase the number of classified elements. The validations/evaluations of the classifications obtained are quite rudimentary, involving a comparison with the initial classification and a manual evaluation of about two hundred items. In some cases it was concluded that, although the classification obtained differs from the original, both are correct.

Waltman et al. (2020) argue that any classification has legitimacy, depending on its intended use, although sometimes it is necessary to establish absolute measures of precision. To this end, they propose that, when comparing two different classifications, a third should be used which is based on a method as different as possible from that of the other two.

Šubelj et al. (2016) analyse multiple automatic classification systems, and value as desirable characteristics the uniformity of the sizes of the thematic areas, that there are no areas which are too small, and other details of automatic classifications such as computing time or stability of the results.

Although with application to a very specific area, Zhang et al. (2022) compare various classifications applied to papers from the multidisciplinary journal Nature, including one based on the authors' labeling of topics. They find major discrepancies between the automated

reference- or content-based classifications and those based on human labeling. They also stress the importance of the assignments to multiple categories.

The WoS assigns categories to scientific journals, with in total about 250 categories. More than one category may be assigned to a given journal and therefore to all of its papers, and there is a Multidisciplinary category. The average number of categories assigned per journal is 1.6 (Wang & Waltman, 2016). Some scientometric studies have grouped these scientific categories into other broader areas, thus forming a hierarchical classification.

In Scopus, an average of 2.1 categories are assigned to each journal (Wang & Waltman, 2016), and the average per paper across the journals rises to 2.50, far more than that of the WoS. There has been a notable increase in recent years (see Fig. 1), indicative of growing interest in classification systems with multiple categories.

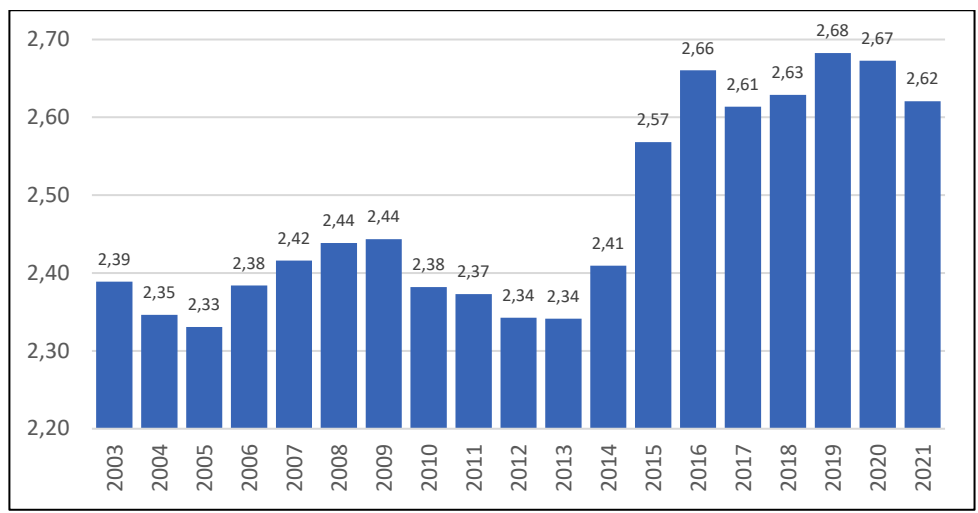

**Fig 1** *Average number of categories per paper in Scopus by year*

In this work, we shall be studying the application of a method to fractionally assign non-multidisciplinary ASJC categories to all the citable papers published in Scopus in 2020, based on the category assignment system that uses the multi-generation parametric method proposed by Glänzel et al. (2021). We shall obtain different classifications by varying some configuration parameters, and analyse the characteristics of each to select those that provide the most homogeneous categories. This selection we shall then compare with the following other two classifications:

a) The "fractional assignment" of the classification of Scopus journals in 2020, in which the Multidisciplinary area and the miscellaneous categories have been eliminated and relocated (Álvarez-Llorente et al., 2023).

b) The "AAC" (Author's Assignation Collection) classification formed by the set of assignments made by the papers' corresponding authors as research guarantors (Moya-Anegón et al., 2013), as compiled by Álvarez-Llorente et al. (2023). In this classification, the papers' own authors fractionally establish the assignments they consider most accurate using the Scopus category scheme (again excluding the Multidisciplinary area and the miscellaneous categories) which covers a subset of the Scopus publications of 2020.

The main purpose of limiting the data sample to the citable papers published in Scopus in 2020 is to allow these comparisons, but it also has the advantage of facilitating the calculations. We consider also that this is a data set that not only quite faithfully represents current citation trends but is also already reasonably well established.

In this comparison we wish to respond to the following research questions:
   a. Which methods coincide most closely with the AAC?
   b. Which methods obtain categories with papers that are most homogeneous in terms of citation habits?
   c. What kinds of papers cannot be classified using these methods, and how many of them are there?
   d. What granularity is obtained with these methods?
   e. How large are the categories obtained?
   f. How are papers classified which are published in journals assigned to the Multidisciplinary category?
   g. Are papers published in miscellaneous categories classified into categories of the same scientific area?
   h. Are classifications obtained that have characteristics more desirable for Scientometrics than those based on ASJC journal assignments?

In the following Method and Data section, we shall explain in detail the proposed classification algorithm with all its parametric variations as well as the initial treatment of the sample data, and identify the works that cannot be classified.

In the Results section, we shall respond to the rest of the research questions by analyzing the classifications obtained with the different parameter combinations. Firstly, the number, size, and granularity of the categories obtained will be examined, as well as the distribution of references among the different categories. This will provide us with a first approximation to the combinations of parameters that provide the scientometrically most desirable results. In a second step, we shall compare the different classifications with the AAC given by the corresponding authors. This will complement the information obtained to allow us to assess whether the proposed classifications seem appropriate and with which parameter combinations they are most accurate so as thus to select those best for further in depth analysis. Then with just these selected classifications we shall analyse how the publications are distributed among the ASJC's thematic areas and how they classify the publications that initially belonged to the Multidisciplinary categories and the miscellaneous areas. The result of all these analyses will finally allow us to determine the most interesting combination of parameters.

### Method and Data

Scopus (Hane, 2004; Pickering, 2004) is a more recent but broader database than WoS (Gómez-Crisóstomo, 2011). It classifies journals using the All Science Journal Classification (ASJC) (Elsevier, 2023). This is a hierarchical classification scheme comprising 311 specific scientific areas or categories grouped into 26 scientific areas, plus a Multidisciplinary scientific area. Furthermore, each of the 26 scientific areas includes a "miscellaneous" category, which if eliminated would leave 285 categories grouped into those 26 scientific areas. A "fractional assignment" is defined (Álvarez-Llorente et al., 2023) in which each area's miscellaneous category assignments are distributed among the rest of that area's categories with a weight divided by the number of categories, and the Multidisciplinary area assignments are distributed among all the categories with a weight of 1/285.

In a proposal to use references to assign categories to papers, Glänzel et al. (1999a, 1999b), only took into account the references from journals classified into just one or two categories, and Milojević's (2020) proposal initially only took into account references from journals assigned to a single category. Glänzel et al. (2021), while not citing so directly, do indicate that they adopt a full-counting method for the assignment classification system. We shall also add fractional assignments (which henceforth we shall call weighted-counting as against full-counting), classifying each assignment by the weight given to the category in the journal, so that if a journal is assigned equally to 3 categories then 1/3 of the weight will be added to the count of each category instead of 1, without a Multidisciplinary area or miscellaneous categories.

For our study, we shall be considering the three weight-allocation models proposed by Glänzel et al. (2021) for normalizing the share of research fields, with the weight of the references being based on the generation number in accordance with three schemes – M1 in which only the first generation references are considered, M2 in which only the second generation references are considered, and M3 in which the first generation references are weighted by 0.618 and the second generation references by 0.382.

Some first generation references may be over-represented with many second generation references, and others under-represented with few second generation references, therefore altering the proportions. For this reason, we introduced a second level of normalization for the second generation by dividing them by the number of them that come from each of the first generation. In this way, the weights of all the second generation references that come from any one of the first generation will sum to unity. Nonetheless, we shall evaluate the impact of this averaging procedure by presenting the results of both applying and not applying it (of course, only in cases where second generation references are used, i.e., in the M2 and M3 schemes).

Milojević (2020) places particular interest in achieving assignments exclusively to a single category. Any tie between categories is broken by opting for the paper's original category derived from the journal. In contrast, Glänzel et al. (2021) use a procedure to assign up to three categories, as long as the second has at least a threshold of two-thirds of the normalized share of the first category and the third has at least two-thirds of the second. We shall be using this system but including up to 5 categories and, instead of always using two-thirds, we shall be testing three different thresholds – 1/2, 2/3, and 4/5.

In sum, we shall be generating a large set of reference-based classifications using the three schemes (M1, M2, M3), with each scheme being combined using the full-counting and the weighted-counting methods to find the share of references, and in turn combined with averaged and non-averaged counting (except for M1 in which second generation references are not counted). The three thresholds will be applied for each of these combinations.

Note that, although the starting point is the fractional ASJC classification (by journals), the proposed classification in whichever of its variants is at the article level since each publication will be classified into one of the ASJC categories (other than Multidisciplinary or miscellaneous) based solely on the references it contains, not on the journal to which it belongs.

None of the reference counting based methods discussed above (Glänzel et al., 1999a and 1999b; Milojević, 2020; Glänzel et al., 2021) allude to any minimum number of active references (i.e., references to which categories can be assigned) for a paper's assignment to be taken as valid. It can therefore be understood that a single active reference is sufficient for a classification to be made.

Table 1 gives the percentages of the year 2020 papers indexed in Scopus with a low number of active references from the first generation, from the second generation, and the sum of the first and second generations. Note that a work may have a low number of active references in the first generation (even none) but a high number in the second, for example, if it cites books (that are not assignable to categories) in which assignable articles are referenced. And just the opposite can also be the case.

*Table 1* *Percentage of 2020 production indexed in Scopus with more than a given number of active references*

| With more than: | % 1st Ref Gen. | % 2nd Ref Gen. | % 1st + 2nd Ref Gen. |
|---|---|---|---|
| 0 | 93.4961 | 93.0944 | 93.6465 |
| 1 | 91.5101 | 92.7908 | 93.2002 |
| 2 | 89.6126 | 92.5382 | 93.1015 |
| 3 | 87.6605 | 92.3164 | 93.0744 |
| 4 | 85.6426 | 92.1062 | 93.0632 |
| 5 | 83.5821 | 91.9129 | 93.0558 |
| 6 | 81.5143 | 91.7258 | 93.0501 |
| 7 | 79.4200 | 91.5454 | 93.0455 |
| 8 | 77.3048 | 91.3731 | 93.0414 |
| 9 | 75.1617 | 91.2047 | 93.0372 |

We believe that an assignment cannot be considered effective if it is not based on at least three active references, so as to be capable of providing a minimum of significance. This means that we shall only be making assignments in 89.61% of the papers with first generation references, 92.53% with second generation references, and 93.10% with both the first and second generation combined. Thus, in all the classifications that we generate, the documents not assigned due to lack of references maintain their initial assignment by journal.

Table 2 presents the analysis by category, and prior to the fractioning of the Multidisciplinary category, of those documents not reassigned due to having fewer than three active references. The Weight columns indicate the weight accumulated by each category (recall that each paper is distributed proportionally among the assigned categories, so that the total of all the categories is equivalent to the number of works). The percentage columns indicate the percentage of papers assigned to the category that are discarded, and the NI columns show the average of the Normalized Impact (normalized citation, item-oriented field-normalized citation score average) (Lundberg, 2007; Waltman et al., 2011) of the said publications. The summary row, which gives the aggregated data of all categories, shows that the highest discard percentage corresponds to considering only the first generation, and the lowest to including both the first and the second generations. One also sees that they are works of low impact.

For the percentages by area, the highest are in Arts and Humanities with 55.35% of the output, followed by Social Sciences with 23.55%, and Nursing with 14.61%. However, these values fall to 38.69%, 15.52%, and 11.83%, respectively, when the two generations are combined.

*Table 2 Scientific production with at most two active references*

| ASJC | Description | 1st Gen. Ref. | | | 2nd Gen. Ref. | | | 1st & 2nd Gen. Ref. | | |
|---|---|---|---|---|---|---|---|---|---|---|
| | | Weight | % | NI | Weight | % | NI | Weight | % | NI |
| 1000 | Multidisciplinary | 3978.5 | 6.51 | 0.25 | 3555.3 | 5.82 | 0.25 | 3489.7 | 5.71 | 0.26 |
| 1100 | Agricultural and Biological Sciences | 7851.7 | 5.21 | 0.27 | 4915.5 | 3.26 | 0.27 | 4363.4 | 2.90 | 0.28 |
| 1200 | Arts and Humanities | 38758.0 | 55.35 | 0.41 | 30522.9 | 43.59 | 0.36 | 27091.2 | 38.69 | 0.34 |
| 1300 | Biochemistry, Genetics and Molecular Biology | 9381.0 | 5.53 | 0.35 | 8206.9 | 4.84 | 0.38 | 8098.0 | 4.77 | 0.39 |
| 1400 | Business, Management and Accounting | 4652.4 | 9.87 | 0.25 | 3182.2 | 6.75 | 0.23 | 2988.2 | 6.34 | 0.23 |
| 1500 | Chemical Engineering | 2881.7 | 4.43 | 0.20 | 1784.6 | 2.74 | 0.18 | 1674.0 | 2.57 | 0.17 |
| 1600 | Chemistry | 3299.3 | 2.58 | 0.27 | 2679.0 | 2.09 | 0.29 | 2577.3 | 2.01 | 0.30 |
| 1700 | Computer Science | 22452.0 | 8.61 | 0.30 | 11995.0 | 4.60 | 0.26 | 10762.1 | 4.13 | 0.26 |
| 1800 | Decision Sciences | 2108.5 | 9.79 | 0.47 | 1191.2 | 5.53 | 0.39 | 1063.6 | 4.94 | 0.38 |
| 1900 | Earth and Planetary Sciences | 10830.4 | 10.12 | 0.42 | 6751.7 | 6.31 | 0.38 | 5985.5 | 5.59 | 0.35 |
| 2000 | Economics, Econometrics and Finance | 2924.3 | 9.39 | 0.31 | 1886.3 | 6.06 | 0.23 | 1705.3 | 5.48 | 0.23 |
| 2100 | Energy | 7155.4 | 9.37 | 0.36 | 4617.6 | 6.05 | 0.27 | 4028.6 | 5.27 | 0.26 |
| 2200 | Engineering | 35187.9 | 10.29 | 0.34 | 21766.2 | 6.36 | 0.29 | 19293.1 | 5.64 | 0.27 |
| 2300 | Environmental Science | 9594.1 | 7.61 | 0.38 | 6281.6 | 4.99 | 0.34 | 5814.2 | 4.61 | 0.33 |
| 2400 | Immunology and Microbiology | 1379.6 | 3.69 | 0.39 | 1125.8 | 3.01 | 0.41 | 1114.2 | 2.98 | 0.41 |
| 2500 | Materials Science | 10818.4 | 6.10 | 0.38 | 6706.3 | 3.78 | 0.35 | 5960.2 | 3.36 | 0.34 |
| 2600 | Mathematics | 9610.9 | 7.67 | 0.37 | 5161.1 | 4.12 | 0.31 | 3853.2 | 3.07 | 0.30 |
| 2700 | Medicine | 73088.3 | 11.54 | 0.50 | 63084.1 | 9.96 | 0.48 | 62219.8 | 9.83 | 0.48 |
| 2800 | Neuroscience | 1447.1 | 3.75 | 0.53 | 1213.7 | 3.14 | 0.58 | 1206.2 | 3.12 | 0.59 |
| 2900 | Nursing | 4068.3 | 14.61 | 0.41 | 3369.1 | 12.10 | 0.38 | 3293.5 | 11.83 | 0.37 |
| 3000 | Pharmacology, Toxicology and Pharmaceutics | 4284.6 | 7.34 | 0.36 | 3177.1 | 5.44 | 0.41 | 3067.7 | 5.25 | 0.42 |
| 3100 | Physics and Astronomy | 19789.4 | 9.45 | 0.53 | 10959.2 | 5.23 | 0.50 | 9606.6 | 4.59 | 0.49 |
| 3200 | Psychology | 3068.8 | 7.40 | 0.37 | 2425.2 | 5.85 | 0.38 | 2281.6 | 5.50 | 0.40 |
| 3300 | Social Sciences | 44838.9 | 23.55 | 0.29 | 32673.3 | 17.16 | 0.25 | 29545.1 | 15.52 | 0.25 |
| 3400 | Veterinary | 818.3 | 4.93 | 0.30 | 589.8 | 3.55 | 0.28 | 563.1 | 3.39 | 0.28 |
| 3500 | Dentistry | 786.4 | 6.25 | 0.44 | 675.5 | 5.37 | 0.44 | 656.9 | 5.22 | 0.43 |
| 3600 | Health Professions | 2123.7 | 10.13 | 0.67 | 1715.0 | 8.18 | 0.70 | 1624.9 | 7.75 | 0.70 |
| | **Total** | **337178** | **10.39** | **0.38** | **242211** | **7.46** | **0.36** | **223927** | **6.90** | **0.36** |

**Results and Discussion**

## Size and uniformity of the categories

Table 3 lists the results for the classifications generated by, let us recall, the combination of the three citation generation schemes (M1, first generation only; M2, second generation only; M3, first and second generations) plus the fractional ASJC classification as a fourth scheme, with the counts from full-counting and weighted-counting averaged (only M2 and M3) and not averaged (identified as AFC, AWC, FC, and WC), and with the three reference thresholds (1/2, 2/3, and 4/5). The following values are presented for each of these resulting classifications:

- Categories: total number of categories with some assignment.
- Max.Cat.: maximum accumulated weight (that of the category with the greatest accumulated weight).
- Min.Cat.: minimum accumulated weight (that of the category with the smallest accumulated weight).
- C.V.: coefficient of variation of the weights of the different categories.
- Granularity: granularity, as defined in Waltman et al. (2020) and Milojević (2020), i.e., the total number of works divided by the sum of the squares of the works (weights) of each category. Thus, the more balanced the categories are, the greater the granularity.

*Table 3* Structure of the classification obtained with various models taking into account only those papers with more than two active references in both the first and the second generations. Included for each scheme, counting method, and threshold are the number of non-empty categories, the size (cumulative weight) of the largest and the smallest category, the coefficient of variation of the size, and the granularity

| Scheme | Counting method | Threshold | Categories | Max. Cat | Min. Cat | C.V. | Granularity |
|---|---|---|---|---|---|---|---|
| M1 | FC | 0.5 | 285 | 98962.2 | 20.15 | 1.19 | 4.086E-05 |
| M1 | FC | 0.67 | 285 | 125714.6 | 6.60 | 1.33 | 3.563E-05 |
| M1 | FC | 0.8 | 285 | 134326.6 | 3.62 | 1.44 | 3.210E-05 |
| M1 | WC | 0.5 | 285 | 131102.2 | 0.61 | 1.37 | 3.402E-05 |
| M1 | WC | 0.67 | 285 | 143842.6 | 0.27 | 1.47 | 3.102E-05 |
| M1 | WC | 0.8 | 285 | 149271.4 | 0.10 | 1.54 | 2.928E-05 |
| M2 | FC | 0.5 | 285 | 109138.6 | 1.70 | 1.38 | 3.394E-05 |
| M2 | FC | 0.67 | 285 | 131357.5 | 0.39 | 1.48 | 3.079E-05 |
| M2 | FC | 0.8 | 285 | 142868.0 | 0.29 | 1.64 | 2.675E-05 |
| M2 | WC | 0.5 | 283 | 142231.7 | 0.08 | 1.52 | 2.951E-05 |
| M2 | WC | 0.67 | 281 | 156330.7 | 0.08 | 1.62 | 2.684E-05 |
| M2 | WC | 0.8 | 280 | 162521.6 | 0.08 | 1.70 | 2.486E-05 |
| M2 | AFC | 0.5 | 285 | 125942.2 | 0.47 | 1.41 | 3.310E-05 |
| M2 | AFC | 0.67 | 285 | 147014.2 | 0.20 | 1.53 | 2.974E-05 |
| M2 | AFC | 0.8 | 285 | 157605.0 | 0.10 | 1.67 | 2.625E-05 |
| M2 | AWC | 0.5 | 282 | 147138.5 | 0.11 | 1.53 | 2.919E-05 |
| M2 | AWC | 0.67 | 281 | 161801.3 | 0.11 | 1.63 | 2.648E-05 |
| M2 | AWC | 0.8 | 280 | 168578.4 | 0.08 | 1.72 | 2.453E-05 |
| M3 | FC | 0.5 | 285 | 108671.2 | 4.09 | 1.31 | 3.630E-05 |
| M3 | FC | 0.67 | 285 | 129450.1 | 1.88 | 1.41 | 3.304E-05 |
| M3 | FC | 0.8 | 285 | 140578.8 | 1.04 | 1.54 | 2.929E-05 |
| M3 | WC | 0.5 | 284 | 138555.0 | 0.27 | 1.44 | 3.187E-05 |
| M3 | WC | 0.67 | 283 | 151483.8 | 0.14 | 1.54 | 2.908E-05 |
| M3 | WC | 0.8 | 282 | 156968.4 | 0.03 | 1.60 | 2.724E-05 |
| M3 | AFC | 0.5 | 285 | 116880.8 | 2.45 | 1.33 | 3.552E-05 |
| M3 | AFC | 0.67 | 285 | 137488.1 | 1.21 | 1.44 | 3.205E-05 |
| M3 | AFC | 0.8 | 285 | 148354.4 | 0.37 | 1.57 | 2.848E-05 |
| M3 | AWC | 0.5 | 284 | 139940.3 | 0.12 | 1.44 | 3.182E-05 |
| M3 | AWC | 0.67 | 282 | 153179.9 | 0.10 | 1.53 | 2.900E-05 |
| M3 | AWC | 0.8 | 282 | 159032.4 | 0.06 | 1.61 | 2.716E-05 |
| ASJC | - | - | 285 | 75837.7 | 273.23 | 0.93 | 5.258E-05 |
| ASJC | - | 0.5 | 285 | 77220.9 | 206.72 | 1.01 | 4.883E-05 |

| Scheme | Counting method | Threshold | Categories | Max. Cat | Min. Cat | C.V. | Granularity |
|---|---|---|---|---|---|---|---|
| ASJC | - | 0.67 | 285 | 77220.9 | 206.72 | 1.01 | 4.885E-05 |
| ASJC | - | 0.8 | 285 | 77206.6 | 206.72 | 1.00 | 4.89E-005 |

As was to be expected, there exists a clear correlation between the granularity and the coefficient of variation. It was calculated to be -0.9923.

As indicated by the data in Table 3, all the classifications obtained have lower granularity than those based on the ASJC, i.e., since the number of categories is similar, the categories are less equal in size. Increasing the threshold also increases the coefficient of variation and reduces the granularity. With regard to the counting methods, full-counting leads to a lower coefficient of variation and greater granularity than weighted-counting. One can also see that, among those which use the second generation, the averaged methods have a greater coefficient of variation and lower granularity, although all the granularity values lie within a quite narrow range of the same order of magnitude of $10^{-5}$. This is comparatively significantly greater than the value of $1.5 \times 10^{-6}$ reported by Milojević (2020) for the WoS reclassification system and that of the WoS itself which is $2.3 \times 10^{-6}$ according to that same author.

The classification is also very important for the normalization of the impact, since this is not comparable between disciplines. Lancho-Barrantes et al. (2010) showed that the disciplines' impact indicators correlate strongly with the average number of references to the documents in the database. One understands the practice of calculating the normalization of the impacts by dividing by the average number of citations in the discipline and year to be well-founded because one assumes that each discipline's documents are more homogeneous among themselves in terms of the citations they receive, and that their citation habits are closely related (Lancho-Barrantes et al., 2010).

Given this foundation, any new classification that we propose should not lead us to categories that are less homogeneous. To evaluate our classifications in this sense, we calculated both the coefficient of variation of the number of references per paper for each category, but weighted by the degree of membership of the paper to the category, and the average of all categories. It would be undesirable if the new classifications were to have a greater average because this would mean that the new categories include documents that are less homogeneous than the original ones.

The main indicators of normalized impact per paper take the complete citation into account, which means that the reference distribution which will correlate most closely is that of those directed to the database (Scopus in this case). Nevertheless, in Table 4 we present not only those directed to Scopus but also the coefficients of variation of the number of total references. We also include the averages of the coefficients of variation of the number of references, both total and those directed to Scopus, but limited to the 2 and 3 preceding years. One observes in the table that the coefficients of variation of the weighted-counting classifications are lower than those of the full-counting classifications. Also, increasing the threshold leads to lower coefficients of variation. The lowest values correspond to the classifications that use the second generation of references. The effect of the averaging in the second generation is barely perceptible, but in most cases the coefficient of variation tends to increase.

*Table 4* *Average of the coefficients of variation of the number of references per paper in each category for each model, indicator, and threshold: ACVS (Average coefficient of variation of the number of references to the Scopus Source Set); ACVS3 (Average coefficient of variation of the number of references to the Scopus Source Set in the previous 3 years); ACVS2 (Average coefficient of variation of the number of references to the Scopus Source Set in the previous 2 years); ACV (Average coefficient of variation of the number of references); ACV3 (Average coefficient of variation of the number of references in the previous 3 years); ACV2 (Average coefficient of variation of the number of references in the previous 2 years)*

| Scheme | Counting method | Threshold | ACVS | ACVS3 | ACVS2 | ACV | ACV3 | ACV2 |
|---|---|---|---|---|---|---|---|---|
| M1 | FC | 0.5 | 0.81 | 1.12 | 1.26 | 0.70 | 1.02 | 1.16 |
| M1 | FC | 0.67 | 0.80 | 1.12 | 1.25 | 0.69 | 1.02 | 1.15 |
| M1 | FC | 0.8 | 0.80 | 1.11 | 1.24 | 0.69 | 1.01 | 1.14 |
| M1 | WC | 0.5 | 0.81 | 1.12 | 1.25 | 0.70 | 1.03 | 1.16 |
| M1 | WC | 0.67 | 0.81 | 1.12 | 1.26 | 0.70 | 1.02 | 1.16 |
| M1 | WC | 0.8 | 0.81 | 1.13 | 1.26 | 0.70 | 1.02 | 1.16 |
| M2 | FC | 0.5 | 0.80 | 1.13 | 1.27 | 0.69 | 1.04 | 1.20 |
| M2 | FC | 0.67 | 0.80 | 1.12 | 1.26 | 0.69 | 1.03 | 1.18 |
| M2 | FC | 0.8 | 0.79 | 1.11 | 1.25 | 0.68 | 1.02 | 1.16 |
| M2 | WC | 0.5 | 0.80 | 1.11 | 1.24 | 0.69 | 1.01 | 1.15 |
| M2 | WC | 0.67 | 0.79 | 1.10 | 1.23 | 0.69 | 1.00 | 1.14 |
| M2 | WC | 0.8 | 0.79 | 1.10 | 1.23 | 0.69 | 1.01 | 1.14 |
| M2 | AFC | 0.5 | 0.81 | 1.14 | 1.27 | 0.70 | 1.04 | 1.18 |
| M2 | AFC | 0.67 | 0.80 | 1.14 | 1.26 | 0.70 | 1.03 | 1.17 |
| M2 | AFC | 0.8 | 0.80 | 1.12 | 1.25 | 0.69 | 1.02 | 1.15 |
| M2 | AWC | 0.5 | 0.80 | 1.11 | 1.24 | 0.70 | 1.01 | 1.14 |
| M2 | AWC | 0.67 | 0.79 | 1.10 | 1.23 | 0.69 | 1.01 | 1.14 |
| M2 | AWC | 0.8 | 0.79 | 1.10 | 1.23 | 0.69 | 1.01 | 1.13 |
| M3 | FC | 0.5 | 0.81 | 1.12 | 1.25 | 0.70 | 1.03 | 1.16 |
| M3 | FC | 0.67 | 0.80 | 1.12 | 1.25 | 0.69 | 1.03 | 1.16 |
| M3 | FC | 0.8 | 0.80 | 1.12 | 1.24 | 0.69 | 1.02 | 1.15 |
| M3 | WC | 0.5 | 0.80 | 1.11 | 1.24 | 0.70 | 1.02 | 1.15 |
| M3 | WC | 0.67 | 0.80 | 1.11 | 1.24 | 0.70 | 1.01 | 1.15 |
| M3 | WC | 0.8 | 0.80 | 1.10 | 1.24 | 0.69 | 1.01 | 1.14 |
| M3 | AFC | 0.5 | 0.81 | 1.13 | 1.26 | 0.70 | 1.03 | 1.17 |
| M3 | AFC | 0.67 | 0.80 | 1.12 | 1.25 | 0.70 | 1.03 | 1.17 |
| M3 | AFC | 0.8 | 0.80 | 1.11 | 1.23 | 0.69 | 1.01 | 1.14 |
| M3 | AWC | 0.5 | 0.80 | 1.11 | 1.24 | 0.70 | 1.02 | 1.15 |
| M3 | AWC | 0.67 | 0.80 | 1.11 | 1.25 | 0.70 | 1.02 | 1.15 |
| M3 | AWC | 0.8 | 0.80 | 1.10 | 1.25 | 0.69 | 1.01 | 1.15 |
| ASJC | - | - | 0.81 | 1.12 | 1.25 | 0.70 | 1.03 | 1.16 |
| ASJC | - | 0.5 | 0.80 | 1.12 | 1.25 | 0.69 | 1.03 | 1.16 |
| ASJC | - | 0.67 | 0.80 | 1.12 | 1.25 | 0.69 | 1.03 | 1.16 |
| ASJC | - | 0.8 | 0.80 | 1.12 | 1.25 | 0.69 | 1.03 | 1.16 |

## Coincidence with the AAC

Table 5 contrasts the classifications generated here with the AAC generated in our previous work (Álvarez-Llorente et al., 2023). The "%Coinc." column presents the percentages of coincidence between the weights assigned by each classification and those assigned by the authors themselves in the AAC, as was defined in the aforementioned work.

"1st AAC cat. rank" indicates the average rank of the assignment in which, in accordance with each classification generated, the winning categories of each AAC paper appear according to their authors. "1st AAC cat. w/o C" is a count of how many of those winning categories according to the authors do not appear among the 5 assigned in the classification generated. I.e., for each AAC paper we take its winning categories according to the AAC itself, and then we look to see whether those categories are among the 5 possible assigned in the classification obtained. If they are, we note the rank (1 to 5) for each in order to calculate the mean in "1st AAC cat. rank". Otherwise, we add it to the "without classification" count ("1st AAC cat. w/o C"). In the two rightmost columns we record just the opposite. We take the winning categories of the classification generated and calculate the average rank in the AAC classification (column "1st class. cat. rank"), or we count it as not among the 5 possible ("1st class. cat. w/o C").

*Table 5 Comparison of classifications with the AAC. Shown are the percentage of coincidence, the average rank that the winning categories in the AAC obtain in the new classifications, the number of winning categories in the AAC that have no mention in the classification, the average rank that the winning categories in each classification have in the AAC, and the number of winning categories of the new classification that are uncategorized in the AAC*

| Scheme | Counting method | Threshold | %Coinc. | 1st AAC cat. rank | 1st AAC cat. w/o C | 1st class. cat. rank | 1st class. cat. w/o C |
|---|---|---|---|---|---|---|---|
| M1 | FC | 0.5 | 30.11 | 1.67 | 9685 | 1.12 | 11409 |
| M1 | FC | 0.67 | 34.15 | 1.57 | 10544 | 1.12 | 11409 |
| M1 | FC | 0.8 | 37.11 | 1.39 | 11940 | 1.12 | 11409 |
| M1 | WC | 0.5 | 39.34 | 1.81 | 9662 | 1.11 | 6004 |
| M1 | WC | 0.67 | 40.43 | 1.56 | 11391 | 1.11 | 6004 |
| M1 | WC | 0.8 | 40.09 | 1.30 | 12877 | 1.11 | 6004 |
| M2 | FC | 0.5 | 28.83 | 1.96 | 9197 | 1.12 | 6779 |
| M2 | FC | 0.67 | 32.36 | 1.87 | 10010 | 1.12 | 6779 |
| M2 | FC | 0.8 | 35.20 | 1.66 | 11560 | 1.12 | 6779 |
| M2 | WC | 0.5 | 37.77 | 1.93 | 9314 | 1.12 | 6113 |
| M2 | WC | 0.67 | 38.68 | 1.67 | 11170 | 1.12 | 6113 |
| M2 | WC | 0.8 | 38.18 | 1.38 | 12866 | 1.12 | 6113 |
| M2 | AFC | 0.5 | 31.45 | 1.96 | 9328 | 1.12 | 6597 |
| M2 | AFC | 0.67 | 34.18 | 1.83 | 10418 | 1.12 | 6597 |
| M2 | AFC | 0.8 | 36.01 | 1.58 | 12041 | 1.12 | 6597 |
| M2 | AWC | 0.5 | 38.21 | 1.90 | 9267 | 1.12 | 6007 |
| M2 | AWC | 0.67 | 39.27 | 1.65 | 11134 | 1.12 | 6007 |
| M2 | AWC | 0.8 | 38.87 | 1.36 | 12771 | 1.12 | 6007 |
| M3 | FC | 0.5 | 30.42 | 1.93 | 8799 | 1.12 | 6425 |
| M3 | FC | 0.67 | 34.10 | 1.83 | 9695 | 1.12 | 6425 |

| Scheme | Counting method | Threshold | %Coinc. | 1st AAC cat. rank | 1st AAC cat. w/o C | 1st class. cat. rank | 1st class. cat. w/o C |
|--------|-----------------|-----------|---------|-------------------|---------------------|----------------------|------------------------|
| M3 | FC | 0.8 | 36.99 | 1.63 | 11231 | 1.12 | 6425 |
| M3 | WC | 0.5 | 39.42 | 1.88 | 9054 | 1.11 | 5907 |
| M3 | WC | 0.67 | 40.43 | 1.63 | 10927 | 1.11 | 5907 |
| M3 | WC | 0.8 | 40.21 | 1.35 | 12521 | 1.11 | 5907 |
| M3 | AFC | 0.5 | 31.75 | 1.94 | 8826 | 1.11 | 6363 |
| M3 | AFC | 0.67 | 35.22 | 1.82 | 9850 | 1.11 | 6363 |
| M3 | AFC | 0.8 | 37.38 | 1.59 | 11539 | 1.11 | 6363 |
| M3 | AWC | 0.5 | 39.52 | 1.87 | 9059 | 1.11 | 5893 |
| M3 | AWC | 0.67 | 40.53 | 1.63 | 10896 | 1.11 | 5893 |
| M3 | AWC | 0.8 | 40.17 | 1.34 | 12569 | 1.11 | 5893 |
| ASJC | - | - | 35.16 | 1.88 | 6662 | 1.19 | 118356 |
| ASJC | - | 0.5 | 37.16 | 1.02 | 9336 | 1.19 | 118356 |
| ASJC | - | 0.67 | 37.15 | 1.02 | 9338 | 1.19 | 118356 |
| ASJC | - | 0.8 | 37.15 | 1.02 | 9340 | 1.19 | 118356 |

A swift overview of the table shows that 0.5 is the threshold with the lowest coincidence percentages, and that the values for the other two are quite similar. Taking only the second generation of references into account yields the poorest results, and weighted-counting gives better results than full-counting. With regard to those using the second generation of references, once again there are no major differences between the averaged and the non-averaged results. What is most striking is that those based on full-counting are below the ASJC, while those based on weighted-counting are above it.

For the "1st AAC cat. rank" and "1st AAC cat. w/o C" columns, one observes that the results are clearly related to the use of the different thresholds – higher thresholds tend to assign fewer categories, thus increasing the count of unassigned winning categories but at the same time lowering the rank of those which are assigned. In the case of the two rightmost columns ("1st class. cat. rank" and "1st class. cat. w/o C"), although there are very few variations, it is intuitive to interpret that the higher the threshold the fewer the categories assigned, and therefore the lower the average rank they are assigned and the fewer of them that remain to be assigned. It is striking that ASJC classifications with a threshold obtain a lower "1st AAC cat. rank", and that they do not have an excessive number of winners in the unassigned AAC ("1st AAC cat. w/o C"). From the opposite perspective however (the "1st class. cat. rank" and "1st class. cat. w/o C" columns), the ranks are significantly higher, and the number of unclassified assignments rises notably.

Table 6 presents the data on how the different methods classify the documents included in the AAC, and how the AAC assignment itself does so – the total number of assigned categories, average number of categories assigned per work, and the percentage of works with 1, 2, 3, 4, and 5 or more assignments (actually, the no-threshold ASJC scheme is the only one that can present more than 5 assignments).

As explained in Álvarez-Llorente et al. (2023), the average number of categories assigned by ASJC

is very high because it repartitions the Multidisciplinary category assignments into low-weight assignments to all categories, and the miscellaneous categories into all of those in the corresponding area. Setting a threshold reduces it by half, although it is still very high.

The AAC has an average of 1.94 categories assigned to each paper. There are only 5 methods which have an average of fewer than 2 categories per paper, and they all have the threshold 0.8 and use weighted-counting. The AAC grants five categories to only 4.68% of the documents, an especially low percentage. The aforementioned 5 methods also assign fewer than 5%, although more than 64% are assigned to a single category compared with 44.85% for the AAC.

*Table 6 Comparison of the assigned categories with the AAC collection*

| Scheme | Counting method | Threshold | Nº assign. | Avg. assign. | %1 | %2 | %3 | %4 | %5+ |
|---|---|---|---|---|---|---|---|---|---|
| M1 | FC | 0.5 | 83302 | 6.59 | 9.85 | 15.03 | 14.26 | 13.34 | 47.53 |
| M1 | FC | 0.67 | 61152 | 4.83 | 24.26 | 19.24 | 13.34 | 10.42 | 32.74 |
| M1 | FC | 0.8 | 39384 | 3.11 | 44.28 | 22.43 | 10.88 | 6.36 | 16.04 |
| M1 | WC | 0.5 | 40309 | 3.19 | 29.83 | 15.17 | 8.43 | 4.94 | 41.64 |
| M1 | WC | 0.67 | 28157 | 2.23 | 48.91 | 20.07 | 9.19 | 4.90 | 16.92 |
| M1 | WC | 0.8 | 20013 | 1.58 | 66.85 | 19.36 | 6.99 | 2.77 | 4.03 |
| M2 | FC | 0.5 | 61030 | 4.72 | 7.23 | 5.06 | 4.27 | 3.50 | 79.93 |
| M2 | FC | 0.67 | 50716 | 3.93 | 20.56 | 9.75 | 5.71 | 3.58 | 60.40 |
| M2 | FC | 0.8 | 36364 | 2.81 | 40.84 | 15.91 | 6.69 | 3.22 | 33.34 |
| M2 | WC | 0.5 | 43367 | 3.36 | 27.59 | 12.63 | 6.67 | 3.10 | 50.01 |
| M2 | WC | 0.67 | 30463 | 2.36 | 46.40 | 18.90 | 8.73 | 4.52 | 21.44 |
| M2 | WC | 0.8 | 21453 | 1.66 | 64.96 | 18.86 | 7.24 | 3.13 | 5.82 |
| M2 | AFC | 0.5 | 55643 | 4.31 | 13.93 | 7.23 | 3.65 | 1.83 | 73.36 |
| M2 | AFC | 0.67 | 44511 | 3.45 | 28.99 | 12.94 | 5.77 | 2.66 | 49.64 |
| M2 | AFC | 0.8 | 31416 | 2.43 | 48.69 | 17.84 | 6.93 | 2.89 | 23.65 |
| M2 | AWC | 0.5 | 43000 | 3.33 | 28.31 | 12.74 | 6.51 | 3.20 | 49.24 |
| M2 | AWC | 0.67 | 30074 | 2.33 | 47.77 | 18.24 | 8.48 | 4.57 | 20.95 |
| M2 | AWC | 0.8 | 21088 | 1.63 | 66.07 | 18.38 | 7.06 | 3.31 | 5.19 |
| M3 | FC | 0.5 | 60940 | 4.70 | 8.16 | 5.31 | 3.38 | 2.41 | 80.74 |
| M3 | FC | 0.67 | 50253 | 3.88 | 22.20 | 10.44 | 5.30 | 2.81 | 59.25 |
| M3 | FC | 0.8 | 36083 | 2.78 | 42.18 | 16.07 | 6.46 | 3.09 | 32.20 |
| M3 | WC | 0.5 | 42375 | 3.27 | 29.08 | 13.21 | 7.24 | 3.53 | 46.95 |
| M3 | WC | 0.67 | 29867 | 2.30 | 47.73 | 19.05 | 8.66 | 4.64 | 19.91 |
| M3 | WC | 0.8 | 21073 | 1.63 | 65.63 | 19.57 | 6.63 | 3.17 | 5.00 |
| M3 | AFC | 0.5 | 58325 | 4.50 | 11.29 | 5.97 | 3.37 | 1.70 | 77.67 |
| M3 | AFC | 0.67 | 46825 | 3.61 | 26.21 | 12.02 | 5.35 | 2.27 | 54.15 |
| M3 | AFC | 0.8 | 33008 | 2.55 | 46.83 | 16.89 | 6.46 | 2.86 | 26.95 |
| M3 | AWC | 0.5 | 42212 | 3.26 | 29.35 | 13.21 | 7.27 | 3.57 | 46.59 |
| M3 | AWC | 0.67 | 29829 | 2.30 | 47.96 | 18.80 | 8.70 | 4.67 | 19.86 |
| M3 | AWC | 0.8 | 20918 | 1.61 | 66.53 | 18.75 | 6.77 | 2.99 | 4.95 |
| ASJC | - | - | 248621 | 18.49 | 15.01 | 18.83 | 11.83 | 7.09 | 47.25 |
| ASJC | - | 0.5 | 132581 | 9.86 | 25.00 | 26.64 | 16.62 | 9.88 | 21.86 |
| ASJC | - | 0.67 | 132572 | 9.86 | 25.04 | 26.62 | 16.61 | 9.88 | 21.85 |

| Scheme | Counting method | Threshold | Nº assign. | Avg. assign. | %1 | %2 | %3 | %4 | %5+ |
|---|---|---|---|---|---|---|---|---|---|
| ASJC | - | 0.8 | 132564 | 9.86 | 25.06 | 26.61 | 16.61 | 9.87 | 21.85 |
| AAC | - | - | 26141 | 1.94 | 44.85 | 30.67 | 14.41 | 5.39 | 4.68 |

For all of the above, we consider that the most promising methods are those that use the 0.8 threshold and weighted-counting since, although they have somewhat lower granularity, they have the greatest coincidence with the AAC (only very slightly surpassed by those with the 2/3 threshold) and the resulting classifications give rise to classes more homogeneous in number of references and to fewer categories assigned to each paper, characteristics that are highly desirable in scientometrics. For this reason, we shall focus on them as methods selected for the following analyses.

### Distribution by category and subject area

In these analyses, we include the AAC classification in the comparisons. Hence, for the other classification methods, we shall necessarily limit their application to the publications included in the said classification (as in Tables 6 and 7). This is not a problem since, as Álvarez-Llorente et al. (2023) demonstrated, the sample of publications included in the AAC classification "represents with great fidelity the thematic variety by areas and categories, as well as by countries of affiliation of their authors" of Scopus publications in 2020.

Table 7 presents the correlations of the resulting sizes of each category (sum of weights) when classifying the AAC documents according to the AAC itself, the 5 selected classification systems, and the fractional AJSC. All are fairly strong correlations. The 5 selected systems have high correlations with each other, slightly lower with the fractional ASJC, and somewhat lower with the AAC, demonstrating that these classifications maintain high coherence with the original ASJC classification.

*Table 7 Correlations of the sum of weights in the ASJC categories between the 5 classification systems with a 0.8 threshold, the AAC, and the fractional ASJC, limited to the documents included in the AAC*

|  | AAC | ASJC | M1 WC 0.8 | M2 WC 0.8 | M2 AWC 0.8 | M3 WC 0.8 | M3 AWC 0.8 |
|---|---|---|---|---|---|---|---|
| AAC | 1 | 0.889 | 0.839 | 0.796 | 0.793 | 0.822 | 0.821 |
| ASJC | 0.89 | 1 | 0.891 | 0.848 | 0.848 | 0.873 | 0.873 |
| M1 WC 0.8 | 0.839 | 0.891 | 1 | 0.985 | 0.985 | 0.997 | 0.997 |
| M2 WC 0.8 | 0.796 | 0.848 | 0.985 | 1 | 0.997 | 0.993 | 0.993 |
| M2 AWC 0.8 | 0.793 | 0.848 | 0.985 | 0.997 | 1 | 0.993 | 0.994 |
| M3 WC 0.8 | 0.822 | 0.873 | 0.997 | 0.993 | 0.993 | 1 | 1.000 |
| M3 AWC 0.8 | 0.821 | 0.873 | 0.997 | 0.993 | 0.994 | 1.000 | 1 |

To delve deeper into analysing the resulting categories in the different classifications, Fig. 2

shows, for each of the classifications, the number of categories that accumulate different weight percentage bands. For example, the band "0.2" indicates how many categories in each classification accumulated a weight percentage in the interval [0.1%, 0.2%). The conclusion that can be drawn from the figure is that the selected classifications show similar distributions among themselves but different from the ASJC and the AAC, being notably greater the number of categories that accumulate less than 0.1%, and also greater although not so markedly the number that accumulate more than 1%. There is a somewhat more even distribution among the intermediate categories.

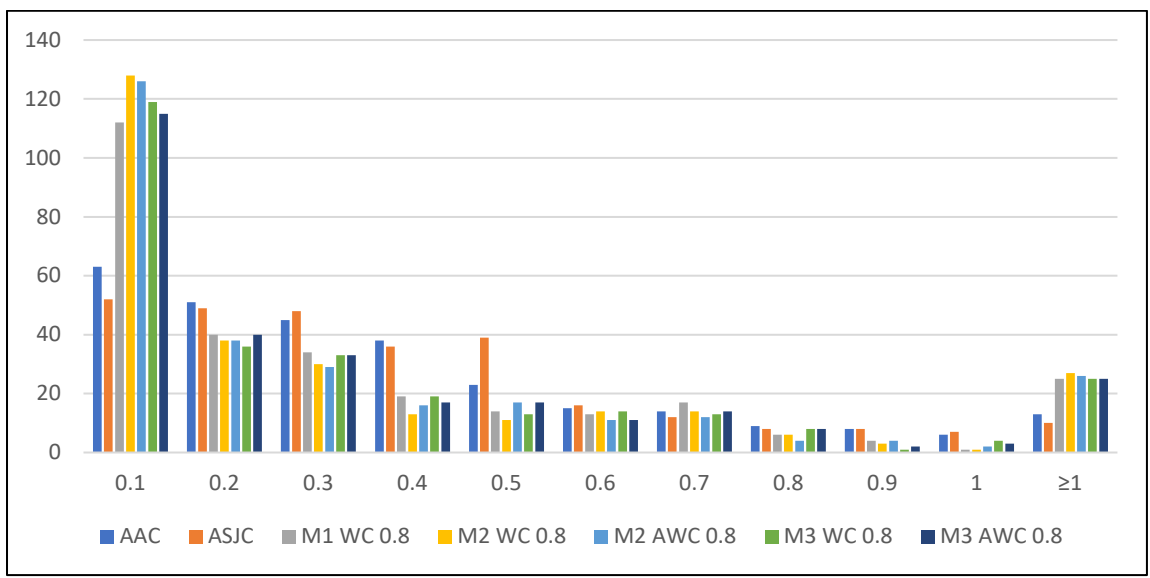

*Fig 2* Number of categories by percentage associated by the 5 classification systems with a 0.8 threshold, the AAC, and the fractional ASJC. Production with 2 or less active references is associated by means of the fractional ASJC

As it is difficult to compare the thematic distribution in the 285 ASJC categories, Table 8 lists the accumulation of weights for each classification, but grouped into the ASJC areas. Fig. 3 illustrates that data graphically. It can be seen that the selected methods once again have a very similar behaviour to each other, and in general also quite similar to that of the ASJC classification. The greatest differences are found in Agricultural and Biological Sciences and in Medicine where there is greater accumulation observed in the selected methods. The opposite is the case for Arts and Humanities, Chemical Engineering, Computer Science, Environmental Science, Materials Science, and Veterinary.

*Table 8* Percentage accumulation (sum of weights) of AAC documents in the 26 ASJC areas according to the 5 classification systems with threshold 0.8, the AAC, and the fractional ASJC

| ASJC | Description | AAC | ASJC | M1 WC 0.8 | M2 WC 0.8 | M2 AWC 0.8 | M3 WC 0.8 | M3 AWC 0.8 |
|---|---|---|---|---|---|---|---|---|
| 1100 | Agricultural and Biological Sciences | 5.23 | 4.90 | 7.21 | 7.31 | 7.31 | 7.30 | 7.26 |
| 1200 | Arts and Humanities | 1.21 | 1.19 | 0.87 | 0.60 | 0.74 | 0.74 | 0.79 |
| 1300 | Biochemistry, Genetics and Molecular Biology | 6.45 | 6.28 | 5.93 | 7.24 | 7.20 | 6.30 | 6.27 |
| 1400 | Business, Management and Accounting | 2.37 | 2.31 | 2.28 | 2.19 | 2.10 | 2.25 | 2.27 |

| ASJC | Description | AAC | ASJC | M1 WC 0.8 | M2 WC 0.8 | M2 AWC 0.8 | M3 WC 0.8 | M3 AWC 0.8 |
|------|-------------|-----|------|-----------|-----------|------------|-----------|------------|
| 1500 | Chemical Engineering | 1.72 | 1.94 | 1.08 | 0.91 | 0.90 | 0.98 | 0.98 |
| 1600 | Chemistry | 3.45 | 3.88 | 3.87 | 4.00 | 3.88 | 3.97 | 3.92 |
| 1700 | Computer Science | 6.80 | 6.11 | 5.37 | 4.93 | 4.93 | 5.09 | 5.12 |
| 1800 | Decision Sciences | 0.77 | 0.57 | 0.40 | 0.36 | 0.35 | 0.38 | 0.36 |
| 1900 | Earth and Planetary Sciences | 3.14 | 3.57 | 4.14 | 4.35 | 4.29 | 4.22 | 4.20 |
| 2000 | Economics, Econometrics and Finance | 2.18 | 1.62 | 2.85 | 3.30 | 3.47 | 3.04 | 3.07 |
| 2100 | Energy | 2.24 | 1.97 | 2.35 | 1.88 | 1.77 | 2.24 | 2.20 |
| 2200 | Engineering | 7.60 | 8.43 | 8.75 | 8.59 | 8.89 | 8.81 | 8.87 |
| 2300 | Environmental Science | 3.94 | 4.45 | 2.60 | 2.23 | 2.14 | 2.43 | 2.44 |
| 2400 | Immunology and Microbiology | 2.21 | 1.42 | 1.77 | 2.01 | 1.78 | 1.83 | 1.79 |
| 2500 | Materials Science | 5.36 | 5.01 | 3.16 | 2.77 | 2.69 | 2.93 | 2.86 |
| 2600 | Mathematics | 5.60 | 4.49 | 4.17 | 3.92 | 4.00 | 4.04 | 4.04 |
| 2700 | Medicine | 17.33 | 19.78 | 21.83 | 22.09 | 22.08 | 21.98 | 21.97 |
| 2800 | Neuroscience | 1.73 | 1.57 | 1.19 | 1.30 | 1.12 | 1.22 | 1.16 |
| 2900 | Nursing | 1.44 | 0.91 | 0.59 | 0.51 | 0.50 | 0.54 | 0.56 |
| 3000 | Pharmacology, Toxicology and Pharmaceutics | 1.65 | 1.91 | 1.76 | 1.50 | 1.47 | 1.63 | 1.63 |
| 3100 | Physics and Astronomy | 4.37 | 6.43 | 6.23 | 6.69 | 6.75 | 6.40 | 6.48 |
| 3200 | Psychology | 2.51 | 2.25 | 2.43 | 2.56 | 2.46 | 2.42 | 2.41 |
| 3300 | Social Sciences | 8.99 | 7.31 | 7.98 | 7.72 | 8.08 | 8.06 | 8.13 |
| 3400 | Veterinary | 0.26 | 0.43 | 0.14 | 0.08 | 0.08 | 0.13 | 0.13 |
| 3500 | Dentistry | 0.34 | 0.45 | 0.51 | 0.51 | 0.52 | 0.52 | 0.53 |
| 3600 | Health Professions | 1.10 | 0.83 | 0.53 | 0.45 | 0.50 | 0.53 | 0.55 |

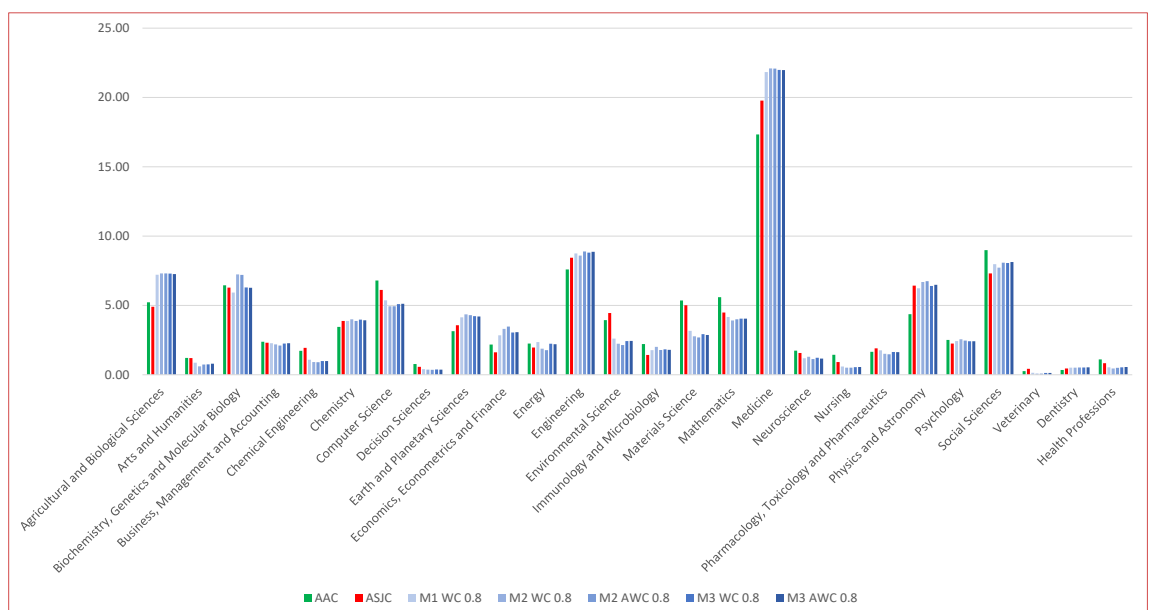

***Fig 3*** *Bar chart representation of the Table 8 data, where the differences between the selected methods and the ASJC classification in certain areas can be appreciated*

## Relocation of multidisciplinary and miscellaneous publications

Possibly one of the most delicate aspects of any attempt at reclassification is how to treat multidisciplinary publications. In what follows therefore we shall analyse how the proposed systems behave with AAC publications that initially belonged to the ASJC Multidisciplinary area (1000) and the miscellaneous categories existing in each of the other areas.

In Table 9 and Fig. 4, one sees how the different strategies classify by area the AAC articles of journals assigned exclusively to the Multidisciplinary category. One can see that some effects commented on for the previous table are repeated, especially the strong increase for the selected weight classifications in Agricultural and Biological Sciences and in Medicine to the detriment of others such as Environmental Science. Nevertheless, in general, despite these striking variations, one sees at a glance that the weights in all strategies conserve a certain correlation, i.e., the distribution of these papers by area is fairly proportional to their size. Indeed, all the columns in Table 9 have a strong correlation with those in Table 8 (greater than 0.8, except for that of AAC which is 0.76).

*Table 9 Distribution of papers published in journals assigned exclusively to the 1000 Multidisciplinary category into the 26 ASJC areas by the 5 classification systems with a threshold of 0.8 and the AAC*

| ASJC | Description | AAC | M1 WC 0.8 | M2 WC 0.8 | M2 AWC 0.8 | M3 WC 0.8 | M3 AWC 0.8 |
|---|---|---|---|---|---|---|---|
| 1100 | Agricultural and Biological Sciences | 8.20 | 15.62 | 15.06 | 15.04 | 15.27 | 15.42 |
| 1200 | Arts and Humanities | 0.87 | 0.50 | 0.30 | 0.30 | 0.36 | 0.36 |
| 1300 | Biochemistry, Genetics and Molecular Biology | 14.64 | 15.10 | 15.42 | 17.54 | 15.49 | 15.42 |
| 1400 | Business, Management and Accounting | 0.39 | 0.42 | 0.35 | 0.41 | 0.41 | 0.41 |
| 1500 | Chemical Engineering | 1.31 | 0.57 | 0.35 | 0.38 | 0.49 | 0.49 |
| 1600 | Chemistry | 1.49 | 2.86 | 1.98 | 2.06 | 2.91 | 2.61 |
| 1700 | Computer Science | 4.04 | 2.87 | 2.76 | 2.46 | 3.00 | 3.17 |
| 1800 | Decision Sciences | 0.18 | 0.18 | 0.08 | 0.14 | 0.08 | 0.08 |
| 1900 | Earth and Planetary Sciences | 4.18 | 5.36 | 6.15 | 6.02 | 5.86 | 5.86 |
| 2000 | Economics, Econometrics and Finance | 3.23 | 3.11 | 3.07 | 3.49 | 2.94 | 3.11 |
| 2100 | Energy | 0.81 | 0.36 | 0.26 | 0.46 | 0.52 | 0.52 |
| 2200 | Engineering | 4.57 | 4.77 | 3.96 | 4.19 | 4.54 | 4.47 |
| 2300 | Environmental Science | 7.18 | 2.77 | 2.63 | 2.04 | 2.54 | 2.57 |
| 2400 | Immunology and Microbiology | 5.37 | 2.44 | 3.66 | 3.06 | 2.88 | 2.81 |
| 2500 | Materials Science | 3.50 | 1.86 | 1.94 | 1.61 | 1.91 | 1.97 |
| 2600 | Mathematics | 3.55 | 1.12 | 0.85 | 0.76 | 1.02 | 0.86 |
| 2700 | Medicine | 13.96 | 19.23 | 20.17 | 19.84 | 19.51 | 19.56 |
| 2800 | Neuroscience | 4.15 | 3.69 | 4.11 | 3.68 | 4.07 | 3.83 |
| 2900 | Nursing | 2.31 | 1.07 | 1.50 | 1.47 | 1.21 | 1.10 |
| 3000 | Pharmacology, Toxicology and Pharmaceutics | 1.71 | 1.56 | 1.15 | 1.24 | 1.38 | 1.61 |
| 3100 | Physics and Astronomy | 3.70 | 4.38 | 5.42 | 4.88 | 4.42 | 4.43 |
| 3200 | Psychology | 2.82 | 3.67 | 2.83 | 3.03 | 3.49 | 3.49 |
| 3300 | Social Sciences | 5.60 | 4.27 | 4.32 | 4.09 | 3.66 | 3.76 |
| 3400 | Veterinary | 0.07 | 0.08 | 0.08 | 0.08 | 0.08 | 0.08 |

| ASJC | Description | AAC | M1 WC 0.8 | M2 WC 0.8 | M2 AWC 0.8 | M3 WC 0.8 | M3 AWC 0.8 |
|---|---|---|---|---|---|---|---|
| 3500 | Dentistry | 0.36 | 1.04 | 0.85 | 0.85 | 0.91 | 0.92 |
| 3600 | Health Professions | 1.80 | 1.12 | 0.74 | 0.87 | 1.05 | 1.10 |

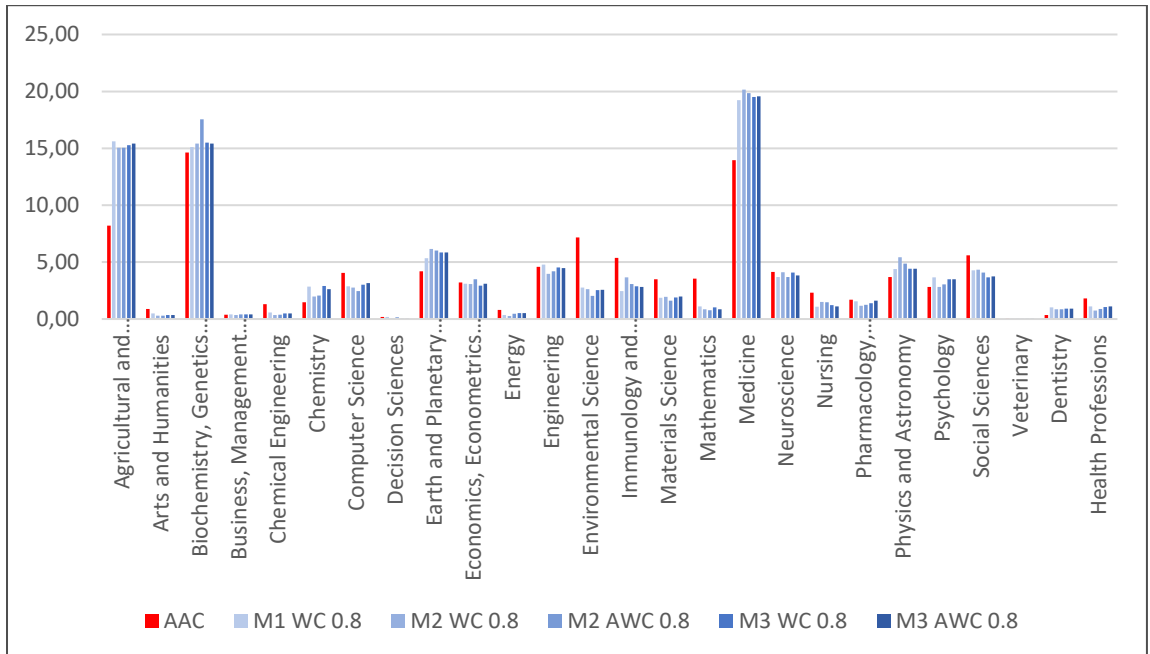

**Fig 4** *Bar chart representation of the Table 9 data, where the greatest differences between the AAC and the selected methods can be appreciated*

In a similar way to how Table 7 presents the correlations between the accumulated weights in the ASJC categories and the different classifications of the AAC publications (Table 6), Table 10 presents those same correlations, but restricted to AAC publications from exclusively the Multidisciplinary category. There is confirmation of major correlations between the proposed classifications, as well as a certain distance between all of them and the AAC.

*Table 10 Correlations between the distributions of papers published in journals assigned exclusively to the 1000 Multidisciplinary category by the 5 classification systems with a 0.8 threshold and the AAC to the 285 ASJC categories*

|  | AAC | M1 WC 0.8 | M2 WC 0.8 | M2 AWC 0.8 | M3 WC 0.8 | M3 AWC 0.8 |
|---|---|---|---|---|---|---|
| AAC | 1 | 0.609 | 0.585 | 0.601 | 0.599 | 0.609 |
| M1 WC 0.8 | 0.609 | 1 | 0.956 | 0.965 | 0.987 | 0.992 |

|            | AAC   | M1 WC 0.8 | M2 WC 0.8 | M2 AWC 0.8 | M3 WC 0.8 | M3 AWC 0.8 |
|------------|-------|-----------|-----------|------------|-----------|------------|
| M2 WC 0.8  | 0.585 | 0.956     | 1         | 0.976      | 0.977     | 0.970      |
| M2 AWC 0.8 | 0.601 | 0.965     | 0.976     | 1          | 0.981     | 0.982      |
| M3 WC 0.8  | 0.599 | 0.987     | 0.977     | 0.981      | 1         | 0.995      |
| M3 AWC 0.8 | 0.609 | 0.992     | 0.970     | 0.982      | 0.995     | 1          |

Finally, in Table 11 we analyse the behaviour in the different classifications of the AAC papers that initially belonged exclusively to the each area's miscellaneous category, indicating the number of papers that belonged to that category and the percentage that stay in that same area for each of the classification methods. In this case, no clear pattern is observed – there are large variations from one area to another, and smaller variations between the different classifications. Most notable is the almost always greater percentage of papers that remain in the same area for the AAC.

*Table 11 Percentage of AAC papers published in journals exclusively assigned to the miscellaneous categories that are assigned to categories belonging to the same ASJC area by the 5 classification systems with a 0.8 threshold and the AAC itself*

| ASJC | Description | Nº | AAC | M1 WC 0.8 | M2 WC 0.8 | M2 AWC 0.8 | M3 WC 0.8 | M3 AWC 0.8 |
|------|-------------|----|-----|-----------|-----------|------------|-----------|------------|
| 1100 | Agricultural and Biological Sciences | 15 | 72.44 | 55.27 | 66.23 | 63.10 | 62.79 | 59.94 |
| 1300 | Biochemistry, Genetics and Molecular Biology | 29 | 72.47 | 77.47 | 74.90 | 78.47 | 77.66 | 77.75 |
| 1400 | Business, Management and Accounting | 9 | 72.78 | 53.84 | 58.99 | 60.42 | 55.56 | 55.56 |
| 1500 | Chemical Engineering | 14 | 65.36 | 46.75 | 35.87 | 29.80 | 39.11 | 43.36 |
| 1600 | Chemistry | 52 | 64.62 | 48.79 | 42.07 | 42.02 | 45.65 | 44.89 |
| 1700 | Computer Science | 63 | 80.61 | 53.41 | 44.33 | 47.80 | 47.83 | 49.01 |
| 1900 | Earth and Planetary Sciences | 51 | 78.73 | 72.12 | 74.82 | 71.91 | 71.68 | 72.84 |
| 2000 | Economics, Econometrics and Finance | 15 | 84.33 | 82.97 | 93.33 | 90.25 | 88.94 | 86.26 |
| 2100 | Energy | 10 | 77.50 | 65.67 | 61.72 | 43.03 | 63.67 | 60.82 |
| 2200 | Engineering | 25 | 55.47 | 55.35 | 48.67 | 52.00 | 48.82 | 48.92 |
| 2300 | Environmental Science | 15 | 53.33 | 28.62 | 25.76 | 21.44 | 27.62 | 28.34 |
| 2500 | Materials Science | 54 | 61.26 | 32.63 | 28.20 | 25.55 | 30.15 | 28.21 |
| 2600 | Mathematics | 89 | 95.00 | 91.95 | 90.83 | 87.97 | 87.60 | 87.21 |
| 2700 | Medicine | 253 | 80.37 | 87.00 | 84.92 | 85.46 | 87.13 | 86.91 |
| 2800 | Neuroscience | 32 | 76.67 | 55.68 | 67.46 | 60.08 | 61.59 | 60.15 |
| 2900 | Nursing | 16 | 78.12 | 37.50 | 29.13 | 27.12 | 31.25 | 31.25 |
| 3000 | Pharmacology, Toxicology and Pharmaceutics | 11 | 58.18 | 31.25 | 18.18 | 26.95 | 25.77 | 33.08 |
| 3100 | Physics and Astronomy | 280 | 37.98 | 51.17 | 46.36 | 46.44 | 46.00 | 46.62 |
| 3200 | Psychology | 37 | 84.50 | 64.94 | 68.08 | 65.67 | 69.68 | 66.71 |
| 3300 | Social Sciences | 12 | 82.08 | 74.54 | 69.03 | 73.59 | 74.50 | 74.55 |
| 3400 | Veterinary | 20 | 59.00 | 21.59 | 14.62 | 16.62 | 25.00 | 22.19 |
| 3500 | Dentistry | 28 | 44.23 | 64.83 | 60.22 | 59.64 | 67.51 | 64.88 |

| ASJC | Description | Nº | AAC | M1 WC 0.8 | M2 WC 0.8 | M2 AWC 0.8 | M3 WC 0.8 | M3 AWC 0.8 |
|---|---|---|---|---|---|---|---|---|
| | Total | 1130 | 66.41 | 63.77 | 60.92 | 60.40 | 61.82 | 61.67 |

**Conclusions**

We have presented a set of methods for classifying papers based on their references in which each publication can be assigned up to 5 weighted categories of the ASJC scheme without the Multidisciplinary area or the miscellaneous categories. The classification methods combine 3 different thresholds for limiting multiple assignments, 4 weighting methods for computing category membership, and 3 schemes of reference generations. To validate them, their scientometric characteristics were compared with the original classification and with the AAC classification done by the papers' authors themselves.

In these classifications, we decided to rule out the reclassification of between about 7% and 11% (depending on the generations used) of publications that did not have a minimum of 3 classifiable references since we considered that in these cases the references provide no significance better than the paper's initial journal-based classification. The allocation is not homogeneous over all thematic areas – that of Arts and Humanities accumulates between 39% and 55%, Social Sciences 16% and 24%, and Nursing 12% and 15%, while the rest accumulate an average of less than 5%. In any case, these are low-impact works, with an average NI of 0.36.

Most of the methods yielded categories that were slightly more homogeneous in citation habits than those of the ASJC, and that had a granularity comparable (slightly lower) with that of the ASJC and an order of magnitude greater than that reported by Milojević (2020) in what was also a reference-based classification but based on the WOS scheme.

Some of the classifications also exceed the ASJC assignment in coincidence with the AAC. What stand out are those that use weighted-counting to compute categories, two generations of references, and parametrized with a threshold of 0.67. Since, however, the results differ little from those with a threshold of 0.8, and with 0.8 the resulting number of categories per paper is closer to what the authors assign in the AAC, this latter threshold was finally selected for the deeper analysis.

There was a strong correlation between the sizes of these classifications and the distribution of weight by subject area with respect to the original journal-based ASJC classification, and a little less with respect to the AAC carried out by the papers' authors.

Analysing how the publications that originally belonged to the Multidisciplinary area are allocated to the different categories showed them to be distributed proportionally to the size of the areas, with the correlation being very high, but much lower with respect to the AAC.

Regarding the reassignment of the publications originally belonging to the miscellaneous categories of each main area, the different methods also behaved very similarly, although they all showed a significantly lower tendency to maintain the assignments within the same principal area than was the case in the AAC classification.

For all these reasons, the classifications parametrized with a 0.8 threshold combined with weighted-counting for the computation of categories demonstrated more desirable characteristics for scientometrics than those derived from the original ASJC journal classification. Among them, we would highlight the one that uses two generations of references and averaged computation (M3-AWC-0.8), which in our opinion, although it is not the best in all areas, shows the most balanced results on average.

However, not much agreement is achieved with the authors' assignments in the AAC. This shows that there is no excessive coherence between the references the authors use in each paper and the categories they themselves assign to it in the AAC. This is in line with the conclusions of Zhang et al. (2022) on categorization based on human labeling, and with Milojević (2020) who indicated that both the original journal-based classification and a reference-based classification are correct.

The application of this classification method has been analyzed for the whole of science, but it can also be applied to specific domains. This would require previously clearly delimiting the scope of the domains as well as having a database with sufficient density of references in those domains.

Lastly, although the study has been constrained to publications registered in Scopus in 2020 with the principal objective of being able to compare its results with the AAC classification of the corresponding authors of our previous work (Álvarez-Llorente et al., 2023), we believe that there would be no obstacle to being able to apply the classification method to any other temporal range or any other database of publications classified by journal.

**Corresponding author**
Correspondence to Vicente P. Guerrero-Bote